\documentstyle[12pt,epsbox]{article} 
\begin{document}
\vspace*{10mm}
\begin{center}
\begin{Large}
{\def\thefootnote{\fnsymbol{footnote}}
\bf Monte-Carlo Simulation of Domain-Wall Network
in Two-dimensional Extended Supersymmetric Theory
\footnote[1]{Talk given at International Symposium on 
Non-Equilibrium 
and Nonlinear Dynamics in Nuclear and Other Finite Systems 
held at May 21-25, 2001 in Beijing.}}\\
\end{Large}
\vspace {7mm}
\begin{large}
Nobuyuki MOTOYUI,
\hspace {10mm}
Shogo TOMINAGA\\
and\\
Mitsuru YAMADA\\
\vspace {7mm}
Department of Mathematical Sciences, Faculty of Sciences,\\
Ibaraki University, Bunkyo 2-1-1, Mito, 310-8512 JAPAN\\
\vspace{5mm}
September, 2001
\end{large}
\end{center}
\vspace{5mm}
\begin{center}
\bf abstract
\end{center}

We will show that 2-dimensional $N=2$-extended supersymmetric
theory can have solitonic solution using the Hamilton-Jacobi method of
classical mechanics. Then it is shown that the Bogomol'nyi mass bound is
saturated by these solutions and triangular mass inequality is satisfied.
At the end, we will mention domain-wall structure in 3-dimensional
spacetime.

\newpage

We will show that 2-dimensional $N=2$-extended supersymmetric
theory can have solitonic solution using the Hamilton-Jacobi method of
classical mechanics. Then it is shown that the Bogomol'nyi mass bound is
saturated by these solutions and triangular mass inequality is satisfied.
At the end, we will mention domain-wall structure in 3-dimensional
spacetime.

The Lagrangean of 2-dimensional $N=2$-extended supersymmetric Wess-Zumino
type model is given as
\begin{equation}
{\cal L}=\int d^{2}\theta d^{2}\theta^{\ast}\phi^{\ast}\phi
+ \int d^{2}\theta W(\phi) + \int d^{2}\theta^{\ast} W(\phi)^{\ast}
\end{equation}
where $\phi$ is a chiral field
\begin{equation}
\phi = a + \sqrt{2}\overline{\theta}^{c}\psi + \overline{\theta}^{c}
\theta f
\end{equation}
and $W(\phi)$ is a superpotential. Where $\psi$ is a 2-dimensional Dirac
spinor
\begin{equation}
\psi = {\psi_{1} \choose \psi_{2}}
\end{equation}
where $\overline{\psi}=\psi^{\dag}\gamma^{0}$ and
$\psi^{c}=C\overline{\psi}^{T}$.
The 2-dimensional $\gamma$-matrices in Majorana representation are
\begin{eqnarray}
\gamma^{0} &=& \sigma_{y}\nonumber\\ 
\gamma^{1} &=& -i\sigma_{x}\nonumber\\
\gamma_{5} &=& \gamma^{0}\gamma^{1} = -\sigma_{z}\nonumber\\
C &=& -\sigma_{y}.
\end{eqnarray}
In component fields, the avobe Lagrangean becomes
\begin{eqnarray}
{\cal L} &=& \partial_{\mu}a^{\ast}\partial^{\mu}a + i\overline{\psi}\gamma
^{\mu}\partial_{\mu}\psi +\frac{i}{2}W''(a)\overline{\psi}^{c}\psi
- \frac{i}{2}W''(a)^{\ast}\overline{\psi}\psi^{c}\nonumber\\
& & - W'(a)W'(a)^{\ast}.
\end{eqnarray}
The current of supersymmetric charge is
\begin{equation}
j_{\mu} = \sqrt{2}\{ \gamma^{\rho}\gamma_{\mu}\psi\partial_{\rho}a^{\ast}
-\gamma_{\mu}\psi^{c}W'(a)^{\ast} \}.  
\end{equation}

After eliminating the fermion field by the fermion field equation of
motion
\begin{equation}
i\gamma^{\mu}\partial_{\mu}\psi - iW''(a)^{\ast}\psi^{c} = 0,   
\end{equation}
we have a purely bosonic Lagrangean
\begin{equation}
{\cal L} = \dot{a}\dot{a}^{\ast} - \nabla a \nabla a^{\ast} - |W'(a)|^{2}.  
\end{equation}

Let us assume that $W(\phi)$ is a polynomial such that
\begin{equation}
W'(a)=0  
\end{equation}
has $n$ complex solution $a_{1},a_{2},\cdots ,a_{n}$.
Then it has not
only $n$ classical vacuum solutions $a(x^{0},x^{1}) = a_{i}$, $i = 1,2,
\cdots ,n$ but also solitonic solutions which we call $(i,j)$-soliton,
characterized by
\begin{eqnarray}
a(t,-\infty) = a_{i},& &
a(t,\infty) = a_{j}.   
\end{eqnarray}

With $(i,j)$-soliton in the background, we can check that the algebra of
supersymmetry charge
\begin{equation}
Q = \int^{\infty}_{-\infty} j^{0}(x) dx 
\end{equation}
undergoes the central extension
\begin{eqnarray}
\left\{ Q, \overline{Q} \right\} &=& 2\gamma_{\mu}P^{\mu}\\
\left\{ Q, \overline{Q}^{c} \right\} &=& -4\gamma_{5} \left[ W(a_{j})
^{\ast} - W(a_{i})^{\ast} \right]. 
\end{eqnarray}
In particular, in the center of mass frame $(P^{\mu}) = (M_{ij},0)$
\begin{eqnarray}
\left\{ Q_{\alpha}, Q_{\beta}^{\dag} \right\} &=& 2M_{ij}\delta_{\alpha
\beta} \\
\left\{ Q_{\alpha}, Q_{\beta} \right\} &=& -4i(\sigma_{x})_{\alpha\beta}
\Delta W^{\ast}  
\end{eqnarray}
where $\Delta W = W(a_{j}) - W(a_{i})$.

From the positivity condition
\begin{equation}
\left\{ A, A^{\dag} \right\} \geq 0  
\end{equation}
with $A = Q_{1} + ie^{i\theta}Q_{2}^{\dag}$, we have
\begin{equation}
M_{ij} \geq 2 {\mbox {\rm Re}} (e^{i\theta} \Delta W).  
\end{equation}
Since $\theta$ is arbitrary, we obtain the lower mass bound of $(i,j)$
soliton
\begin{equation}
M_{ij} \geq 2|\Delta W|.  
\end{equation}
Actually, the Bogomol'nyi bound is saturated by classical solution. To
see this, we calculate the static solution of the field equation. From the
bosonic Lagrangean,
the Hamiltonian is
\begin{equation}
{\cal H} = \dot{a}\dot{a}^{\ast} + (\nabla a)(\nabla a^{\ast}) + |W'(a)|
^{2}
\end{equation}
where $\nabla a = da \slash dx$.
Writing the static solution of $a(t,x)$ simply as $a(x)$, and regarding
$x$ as time,
\begin{eqnarray}
{\cal L}' &=& (\nabla a)(\nabla a^{\ast}) +  |W'(a)|^{2}\\
{\cal H}' &=& p_{a}p_{a}^{\ast} - |W'(a)|^{2}   
\end{eqnarray}
where $p_{a}$ is the conjugate momentum to $a$. This is a problem of one
particle moving in the potential
\begin{equation}
U = -|W'(a)|^{2}.  
\end{equation}

The Hamilton-Jacobi equation for the action $S(a,a^{\ast})$ is
\begin{equation}
\left( \frac{\partial S}{\partial a^{\ast}} \right) \left( \frac{\partial S}
{\partial a} \right) - W'(a)W'(a)^{\ast} = E.
\end{equation}
For $E=0$ we can write the complete solution
\begin{equation}
S(a,a^{\ast},\alpha) = \alpha W(a) + \frac{1}{\alpha}W^{\ast}  
\end{equation}
where $\alpha = e^{i\omega}$ is a parameter. The soliton path is given by
\begin{equation}
\frac{\partial S}{\partial \alpha} = W(a) - \frac{1}{\alpha^{2}}W(a)
^{\ast} = const.  
\end{equation}
Then
\begin{equation}
{\mbox {\rm Im}} (e^{i\omega}W(a)) = const.  
\end{equation}
\begin{figure}[h]
\begin{tabular}{cc}
\begin{minipage}{0.50\hsize}
\psbox[height=60mm]{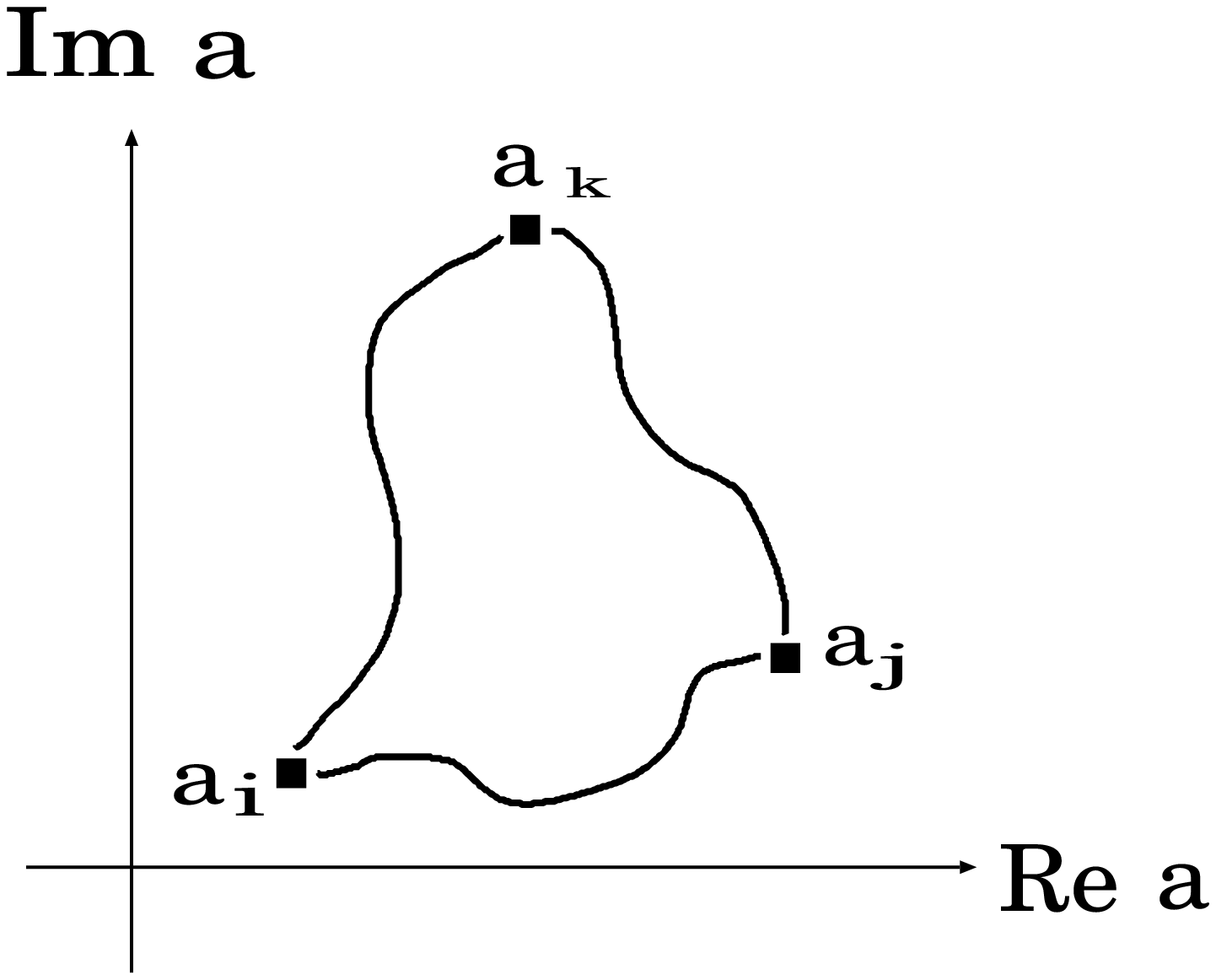}
\caption{Soliton paths in a-plane.}
\label{fig-a}
\end{minipage}&
\begin{minipage}{0.50\hsize}
\psbox[height=60mm]{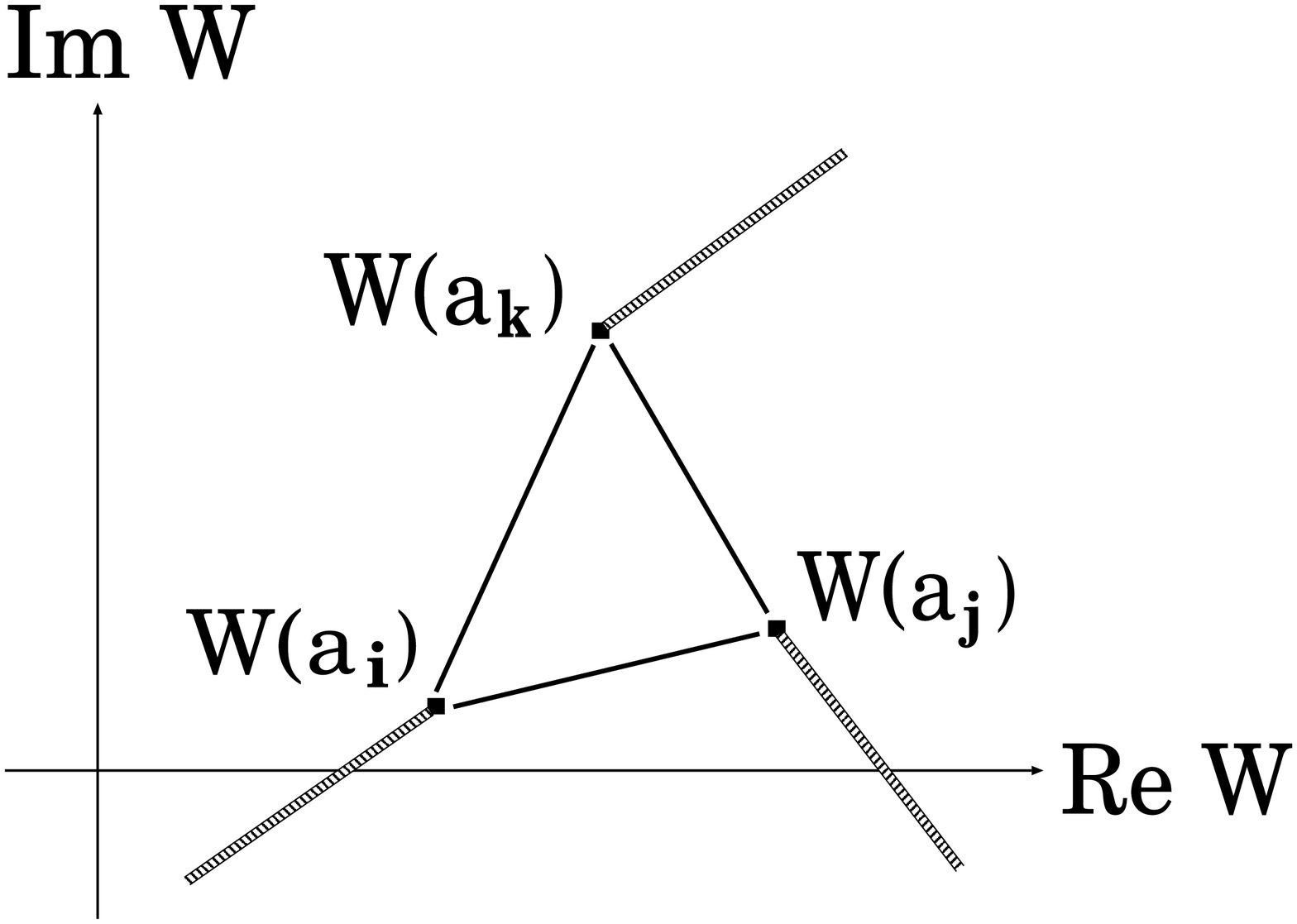}
\caption{Soliton paths in $W$-plane.}
\label{fig-w}
\end{minipage}
\end{tabular}
\end{figure}
So the trajectory of $a(x)$ is such that $W(a(x))$ is a straight line in
complex $W$-plane. In complex $W$-plane, there is a branch cut starts
from each $W(a_{i})$ to infinity because 
\begin{equation}
\left.\frac{da}{dW}\right|_{W=W(a_{i})}=\infty.
\end{equation}
Therefore only solitons whose paths do not cross branch cuts can exist.

The classical mass $M_{ij}$ is obtaind as follows. From $E=0$, we have
\begin{equation}
|\nabla a| = |W'(a)|  
\end{equation}
and
\begin{equation}
dx = \frac{|da|^{2}}{|dW|}  
\end{equation}
so the mass is given as
\begin{eqnarray}
M_{ij} &=& \int^{\infty}_{-\infty} {\cal L'} dx \nonumber\\
&=& 2\int^{\infty}_{-\infty} |W'(a(x))|^{2} dx \nonumber\\
&=& 2 \int |dW|.
\end{eqnarray}
Since $W(a)$ is a straight line in the complex $W$-plane,
\begin{equation}
M_{ij} = 2\left| \int^{W(a_{j})}_{W(a_{i})} dW \right|
= 2|\Delta W|.  
\end{equation}
So the Bogomol'nyi bound is saturated by classical solution.

Then, from the triangular inequality in the complex $W$-plane,
a strict mass inequality\\
\begin{equation}
M_{ik} < M_{ij} + M_{jk}
\end{equation}
follows. This shows the absolute stability of one-soliton configuration
through the attractive force between neighboring solitons.

Now for the general cases of $D$-dimensions $(D=2,3,4)$ the energy of the
system is
\begin{eqnarray}
E &=& \int d^{D-1}x {\cal L}'\\
{\cal L}' &=& \sum^{D-1}_{i=1} |\nabla_{i}a|^{2} + |W'(a)|^{2}. 
\end{eqnarray}

Let us assume that there are $n$ vacuum configurations characterized by
\begin{eqnarray}
W'(a_{i}) = 0 &;& i = 1,2,\cdots ,n. 
\end{eqnarray}

In 3-dimensions, we can summarize the features of low energy
configurations as follows:
\begin{itemize}
\item The 2-dimensional space is divided into\\
$(i)$-domains; $i = 1,2,\cdots ,n$.
\item Every two domains are separated by\\
$(i,j)$-wall; Every domain-wall is a curve in $xy$ space. 
\item $(i,j)$-, $(j,k)$- and $(k,i)$-wall can join at\\
$(i,j,k)$-wall-junction; Every wall-junction is a point in
$xy$ space.
\end{itemize}
We will show some pictures of 2-dimensional networks generated by
Monte-Carlo simulation, where we put 
\begin{equation}
W(\phi) = \frac{1}{4}\phi^{4} - \phi.  
\end{equation}
There are three distinct vacuum configurations
$1,\omega,\omega^{\ast}$
as the solution of
\begin{equation}
W'(a) = a^{3} - 1 = 0 .
\end{equation}

In the Monte-Carlo iterations, field configurations are generated by
the statistical weight of
$e^{-\frac{E}{kT}}$.  
These domain wall configurations are metastable and as iteration goes,
domains tend to be unified into the real vacuum which consists of
single domain.

The auther (N. M.) acknowledges Ibaraki university international academic
exchange grant.
\begin{figure}[tb]
  \begin{tabular}{cc} 
   \begin{minipage}{70mm}
    \psbox[height=70mm]{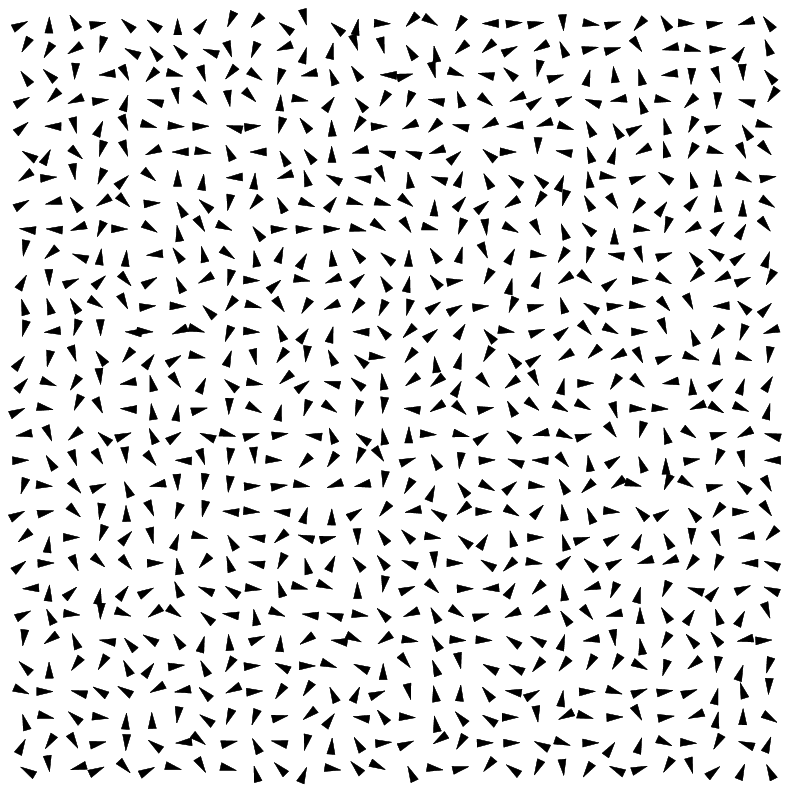}
   \end{minipage}&
   \begin{minipage}{70mm}
    \psbox[height=70mm]{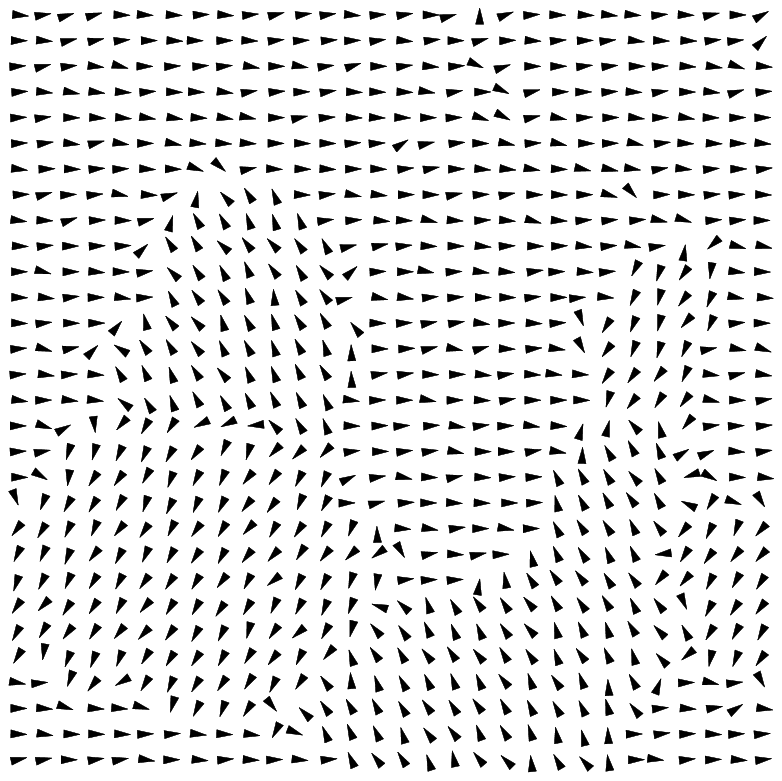}
   \end{minipage}\\
     \vspace{-6mm}
        \\
   \begin{minipage}{70mm}
   \center{(a)}
   \end{minipage}&
   \begin{minipage}{70mm}
   \center{(b)}
   \end{minipage}
  \end{tabular}
  \caption{a) Initial configuration of $a(x,y)$. The length of the
           arrows is normalized.
           b) The configuration after Monte-Carlo iteration.}
  \label{fig-monte}
\end{figure}

\end{document}